\documentclass[%
superscriptaddress,
preprint,
 amsmath,amssymb,
 aps,
]{revtex4-1}

\usepackage{graphicx}
\usepackage{bm}

\usepackage{titlesec}
\titleformat{\section}[block]{\bfseries\centering}{\thesection.}{1em}{\MakeUppercase}

\usepackage[draft]{todonotes} 
\begin{document}

\title{Instability to a heterogeneous oscillatory state in randomly connected recurrent networks with delayed interactions}

\author{C\'{e}lian Bimbard}
\affiliation{Laboratoire des Syst\`{e}mes Perceptifs, \'{E}quipe Audition, CNRS UMR 8248, \'{E}cole Normale Sup\'{e}rieure, Paris, France.}
\author{Erwan Ledoux}
\affiliation{Group for Neural Theory, Laboratoire de Neurosciences Cognitives, INSERM U960, \'{E}cole Normale Sup\'{e}rieure, Paris, France.}
\author{Srdjan Ostojic}
\affiliation{Group for Neural Theory, Laboratoire de Neurosciences Cognitives, INSERM U960, \'{E}cole Normale Sup\'{e}rieure, Paris, France.}

\date{\today}

\begin{abstract}
Oscillatory  dynamics are ubiquitous  in biological  networks. Possible
sources  of  oscillations   are  well  understood  in  low-dimensional
systems,  but  have  not   been  fully  explored  in  high-dimensional
networks. Here we study  large networks consisting of randomly coupled
rate  units.  We identify  a  novel type  of  bifurcation  in which  a
continuous  part  of the eigenvalue spectrum of the linear stability matrix crosses  the  instability line  at
non-zero-frequency. This bifurcation  occurs when the interactions are
delayed  and partially  anti-symmetric, and  leads to  a heterogeneous
oscillatory state in  which oscillations are apparent in the activity of
individual units, but not on the population-average level.
\end{abstract}

\maketitle

\section{Introduction}
Networks of interacting units are a fundamental model of many physical
and  biological   systems,  and  understanding   their  dynamical repertoire  is of  outmost importance.  Dynamical systems theory
  has examined extensively the
dynamics  of low-dimensional systems,  and in  particular bifurcations
between  different  regimes \cite{strogatz}.  These bifurcations  typically
occur when  an eigenvalue of the  system (or pair  thereof) crosses an
instability line,  and enumerating the  possible scenarios leads  to an
exhaustive  taxonomy   of  dynamical  behaviors   for  low-dimensional
networks.

Many biological systems, in particular regulatory and neural networks,
are  however high-dimensional  as they  consist of  a large  number of
individual  units, and   the  interactions  between  units  are
moreover  strongly  disordered.   In contrast to low-dimensional systems, the   dynamical  repertoire  of  high-dimensional
networks  of  randomly interacting  units  has not been fully charted, and new phenomena have been recently discovered.
In particular, within  the context  of randomly  connected neural  networks,  a novel
bifurcation has  been identified, in which a  continuum of eigenvalues
loses stability,  leading to  a transition from  constant, fixed-point
activity to  highly heterogeneous, chaotic  activity \cite{Sompolinsky}.  This transition
and its  implications for neural computations have  lately attracted a
significant amount of attention \cite{Sussillo,Buonomano,Wainrib,Toyoizumi,Ostojic,Kadmon}.

In the bifurcation to chaotic activity described above, the continuous
part of the spectrum that  loses stability is centered around the real
axis  in   the  complex  plane,   and  leads  to  an   instability  at
zero-frequency. Here we show that an analogous, but non-zero frequency
instability  can occur  when  interactions are  delayed and  partially
anti-symmetric. This novel type of  bifurcation leads to a new type of
heterogeneous oscillatory state, in which different units oscillate at
similar frequencies but random phases.

\section{Network model}

We investigated the dynamics of a network of $N$ randomly connected rate units \cite{dayan-abbott} given by

\begin{equation}
	\frac{dx_{i}(t)}{dt} = -x_{i}(t) + \sum_{j=1}^NJ_{ij}\phi(x_{j}(t-D)) \,\,\,\,\,\,i=1\ldots N
	\label{ref_eq}
\end{equation}

where  $x_i$ is  the activity  of unit  $i$, $\phi(x)=\tanh(x)$  is the
transfer  function, and  $D$  represents a  delay  in the  interaction
between the  units.  The  elements of the  interaction matrix  $J$ are
drawn  from   a  Gaussian  distribution  of  mean   $0$  and  variance
$g^2/N$. The symmetric elements $J_{ij}$ and  $J_{ji}$ are correlated,  and the
degree  of symmetry  $\tau_s=\langle J_{ij}J_{ji}\rangle N/g^2$  is a
parameter that we systematically  varied. For $\tau_s=1$, the interaction
matrix is fully  symmetric, for $\tau_s=0$ it is  fully asymmetric, and
for $\tau_s=-1$ it is fully anti-symmetric.

The network possesses a trivial  fixed point for which the activity of
all the units  vanishes. We examined the  stability of this fixed
point as function  of the strength of coupling $g$, the degree  of symmetry in the
interactions  $\tau_s$, and  the delay $D$.  We  considered  heterogeneous perturbations  in
which the activity of each unit is perturbed away from the fixed point
by  an amount  $\delta {x_i}(t)  =\delta{\tilde{x}_i}  e^{\lambda t}$,
with $\lambda$ an arbitrary complex number representing the decay of the perturbation.

Inserting the perturbation in Eq.~\ref{ref_eq} and linearizing around the fixed point we get

\begin{equation}
{e^{\lambda D}}(1+\lambda)\delta \tilde{x}_i =\sum_{j=1}^N {J_{ij}}\delta {\tilde{x}_j}.
\end{equation}

Diagonalizing the interaction matrix $J$, and projecting on the $k$-th eigenmode yields

\begin{equation}
{e^{\lambda_k D}}(1+\lambda_k)\ = \mu_k   \label{characteristic-eq}
\end{equation}
where $\mu_k$ is the $k$-th eigenvalue of $J$. We therefore obtain $N$ independent
characteristic  equations  that   specify $N$ admissible  values  of  the
perturbation decay  $\lambda_k$,   $k=1\ldots N$, one for  each eigenvalue of $J$.  For a
given set of parameters $g, \tau_s$ and $D$, if any of these equations
admits   a   solution   with   $\Re{\lambda_k}>0$,   the   corresponding
perturbation is amplified and the fixed point is unstable.

\begin{figure}
  \includegraphics[width=0.8\textwidth]{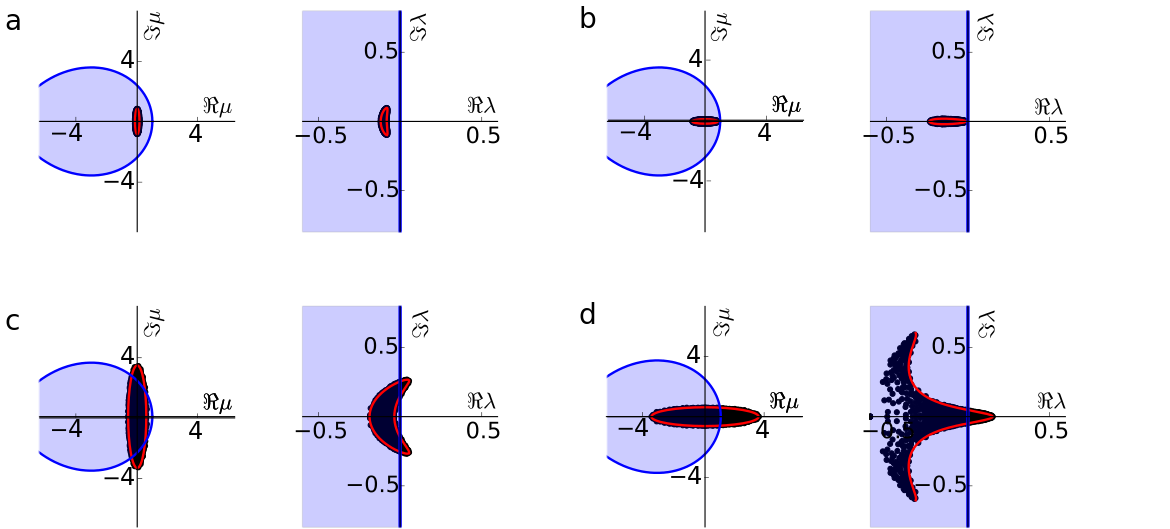}	
  \caption{Linear stability of the fixed point solution.
  a-d. Left panel:  Eigenvalues $\mu$ of a simulated interaction matrix (black dots),
   theoretically-determined domain containing the eigenvalues (red ellipse) and  stability domain (blue shaded region).
  Right panel: perturbation decay $\lambda$ associated to each eigenvalue $\mu$ (black dots), 
  theoretical prediction of the  $\lambda$ associated to the contour of the $\mu$-ellipse (red line) and stability domain (blue shaded region). Parameter values
  a. $\tau_s = -0.7$ and $g = 0.2$ (N = 1000, $D = 0.2$)
  b. Same as in a., with $\tau_s = 0.7$ and $g = 0.2$
  c. Same as in a., with $\tau_s = -0.7$ and $g = 2.0$
  d. Same as in a., with $\tau_s = 0.7$ and $g = 2.0$.
  }
  \label{EVSpectra}
  \end{figure}

In the limit  of large $N$, the eigenvalues $\mu_k$  of the matrix $J$
are uniformly distributed within an  ellipse centered at the origin in
the  complex plane,  with real  and imaginary  axes  $g(1+\tau_s)$ and
$g(1-\tau_s)$                                              respectively
\cite{Sommers}. Solving Eq.~\ref{characteristic-eq} for $\lambda$ yields a mapping that transforms this ellipse into
a different  domain in the  complex plane, that contains  the solutions
$\lambda_k$ of Eq.~\ref{characteristic-eq}  (Fig.~\ref{EVSpectra}).  We   will  call
$\mu$-domain the domain containing the eigenvalues $\mu_k$ of $J$, and
$\lambda$-domain  the domain containing  the solutions  $\lambda_k$ of
Eq.~\ref{characteristic-eq}. Note  that the $\mu$-domain  depends only
on the values of $g$  and $\tau_s$, while the $\lambda$-domain depends
also on the interaction delay $D$.

\section{Phase Diagram}

One  approach for assessing  the stability  of the  fixed point  for a
given  set   of  parameters  is  to  examine   whether  the  resulting
$\lambda$-domain  intersects  the  line  $\Re{z}=0$ in  the  complex
plane. An alternative approach is to use the inverse mapping and determine the contour $\mu(z)$ in
the $\mu$-plane that  corresponds to $\Re{\lambda}=0$, and examine
whether  this  contour  intersects  the  $\mu$-domain  containing  the
eigenvalues.   

The two   approaches   are  graphically   illustrated   in
Fig.~\ref{EVSpectra}. For small values  of the coupling $g$, the  eigenvalues $\mu_k$ of $J$
have   a   small   modulus,    and   all   $\lambda_k$   solution   of
Eq.~\ref{characteristic-eq}  have a negative  real part:  the synaptic
coupling is too low to drive any instability of the fixed point.  When
the  coupling  $g$ is  increased,  the  $\lambda$  domain crosses  the
imaginary axis  (fig. \ref{EVSpectra}-c,d).  In this  case, the degree
of symmetry  of the matrix appears  to play an important  role for the
shape   of   the   $\lambda$-domain:   in   the   antisymmetric   case
(fig. \ref{EVSpectra}-c), values with  $\Re\lambda >0$ have a non-zero
imaginary     part,     whereas      in     the     symmetric     case
(fig. \ref{EVSpectra}-d),  the solution $\lambda_0$  with largest real
part has a  vanishing imaginary part. We predict  that these two cases
are  unstable,   the  antisymmetric   case  possibly  leading   to  an
oscillatory behavior.

To systematically identify  bifurcation points,  we determined the  coupling strength
$g_c$  for which  $\lambda=i\omega_c$ is  a solution  of  a characteristic
equation   Eq.~\ref{characteristic-eq}.    We   first   consider   two
simplifying cases, before describing the general solution.

For a symmetrically skewed matrix ($0  < \tau_s < 1$), the ellipse that
contains the eigenvalues of the matrix $J$ is elongated along the real
axis.  The  eigenvalue $\mu_0$ with  the largest modulus  is therefore
purely  real,  and given  by  $\mu_0=g(1+\tau_s)$.  The  corresponding
solution $\lambda_0$ of Eq.~\ref{characteristic-eq} is also real, and
vanishes   when  $\mu_0=1$,   which  yields   the   critical  coupling
$g_c=1/(1+\tau_s)$.   At  this  coupling,   all  other   solutions  of
Eq.~\ref{characteristic-eq} have negative real parts, hence we have a zero-frequency bifurcation.

For an anti-symmetrically skewed matrix ($-1 < \tau_s < 0$), the ellipse that
contains the eigenvalues of the matrix $J$ is elongated along the imaginary
axis, and  the  eigenvalue $\mu_0$ with  the largest modulus  is therefore
purely  imaginary and given by $\mu_0=ig(1-\tau_s)$. The corresponding solution $\lambda_0$ of 
Eq.~\ref{characteristic-eq}  now has a non-zero imaginary part, and the solution $\lambda_0=i\omega_c$ is given by

\begin{eqnarray}
\omega_c &=& \sqrt{(g(1-\tau_s))	^2-1}  \label{eq_D1}\\
D &=& \frac{1}{\sqrt{(g(1-\tau_s))^2-1}}\arcsin(\frac{1}{g(1-\tau_s)}).  \label{eq_D2}
\end{eqnarray}

This solution leads to a  bifurcation with a non-zero frequency.

More generally,  the first  eigenvalue  that
becomes unstable as the coupling  $g$ is increased does not need to correspond
to an extremity of the ellipse in the $\mu$-plane. To determine the location of the instability
in         the        general         case,         we        followed
\cite{Marcus,Belair},  and determined  the contour
$\mu(z)$  in   the  complex   plane  that  is   mapped  to   the  line
$\lambda=i\omega$ via  Eq.~\ref{characteristic-eq}. Parametrizing this
contour in polar coordinates as $\mu(\theta)=M(\theta)e^{i\theta}$, we have 
for $0 < \theta < \pi$ (the contour is symmetric to the horizontal axis)\cite{Marcus}

\begin{eqnarray}
  M(\theta) &=& \sqrt{1+\omega^2} \nonumber \\
  -\omega &=& \tan(\omega D - \theta), \,\,\,\,\,	\mathrm{with} \,\,\,\, \theta - \frac{\pi}{2} < \omega D < \theta, \,\,\,\,\,	\mathrm{modulo} \,\,\,\, 2\pi .
  \label{def_contour}
\end{eqnarray}

This contour depends only on the value of the delay $D$, and
defines a droplet-shaped stability boundary for  the eigenvalues of $J$ (Fig.~\ref{EVSpectra}). The critical
coupling  $g_c$  is  then  determined  by the  first  intersection  as
coupling is increased between the elliptical $\mu$-domain containing the eigenvalues, and the stability boundary defined by Eq.~\ref{def_contour}.

\begin{figure}
    \includegraphics[width=1.0\textwidth]{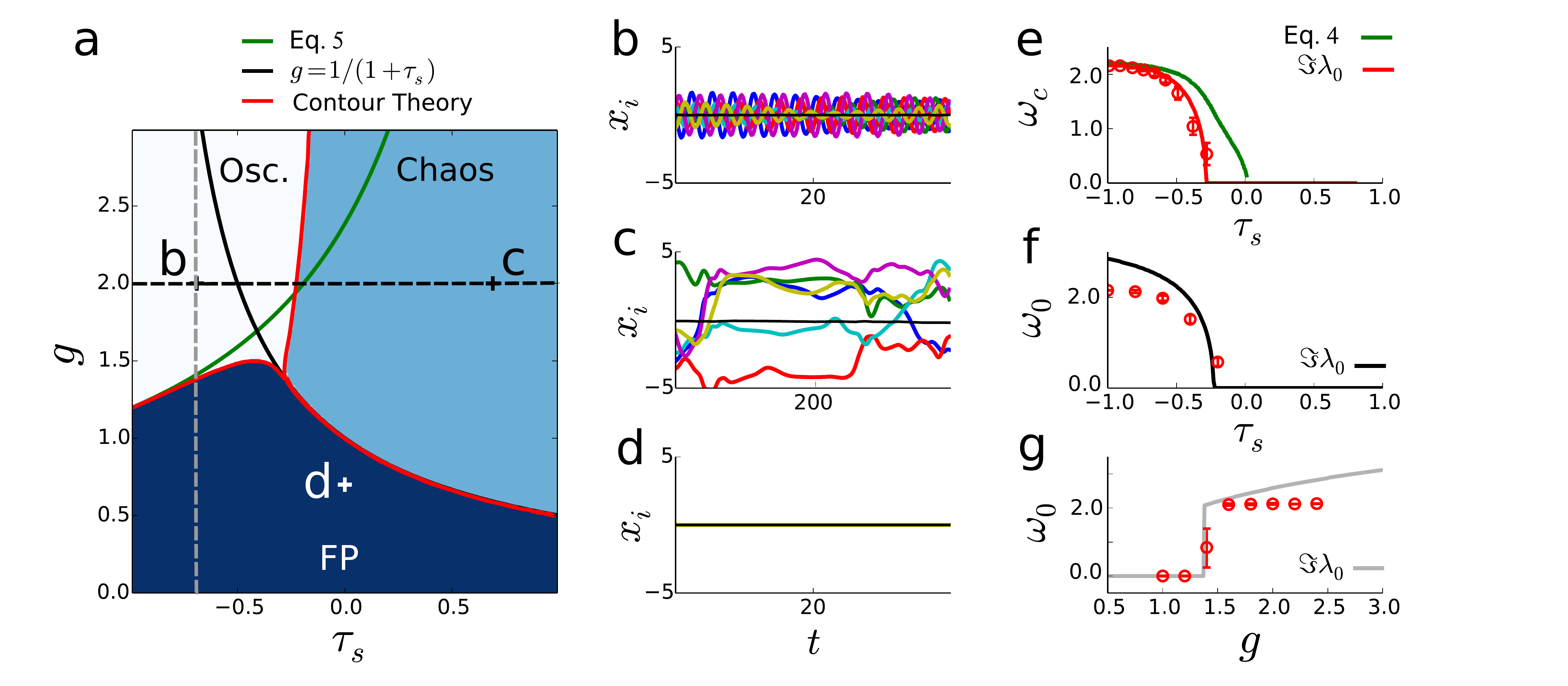}	
    \caption{Dynamical regimes of the network.
    a. Bifurcation Diagram. Dark blue region: fixed point (FP) is stable; 
    light blue region: FP is unstable, with dominant eigenvalue real;
    white region: FP is unstable, dominant eigenvalue with a non-zero imaginary part.
    Red line: exact theoretical prediction for the instability line.
    Black line: approximation for $\tau_s > 0$, 
    Green line: approximation for $\tau_s < 0$.
    b-d: Illustrations of simulated dynamics in different regions of the bifurcation diagram (N = 1000, $D = 0.2$). The panels show 
 firing rates of ten neurons and the  population average in  black. Parameter values:   
    b.  $\tau_s = -0.7$ and $g = 2.0$,
    c.  $\tau_s = 0.7$ and $g = 2.0$,
    d.  $\tau_s = 0.7$ and $g = 0.7$.
    e. Bifurcation frequency $\omega_c$ computed along the instability line in panel a (red line), and approximation given by $\omega_0 = \sqrt{(g(1-\tau_s))^2-1}$ (green line).
    Simulation results are shown as red dots (N = 500, n = 5, mean $\pm$ sem).
    f. Frequency of the dominant eigenvalue for constant $g$ as function of $\tau_s$ (along the black dashed line in panel a), compared to the frequency of oscillations in the simulated network (red dots). g. As in f, but for fixed $\tau_s$ and varying $g$ (along the grey dashed line in panel a). }
    \label{BifDiag}
    \end{figure}

The  full bifurcation  diagram as  function of  symmetry  $\tau_s$ and
coupling  $g$ for  a fixed  value  of the  delay $D$  is displayed  in
Fig.~\ref{BifDiag}.    For  the   symmetric   case,  $\tau_s>0$,   the
instability  is given  by the  extremity of  the ellipse  on  the real
axis. It therefore occurs at zero frequency, and the critical coupling
is given  by $g_c=1/(1+\tau_s)$. For couplings larger  than $g_c$, the
fixed  point is  unstable, and  a chaotic,  fluctuating  state appears
\cite{Sompolinsky}.

The zero-frequency, chaotic  instability extends to the anti-symmetric
region until  a critical value $\tau_s^{c}<0$ for  which a non-zero
frequency bifurcation appears. As $\tau_s$ is further decreased,
the non-zero frequency instability is determined by eigenvalues closer and closer to
the extremity  of the  ellipse along the  imaginary axes, so  that the
critical coupling  and the frequency  of the instability are  well approximated by
Eqs.~\ref{eq_D1}-\ref{eq_D2}.

\begin{figure}[h!]
  \includegraphics[width=1.0\textwidth]{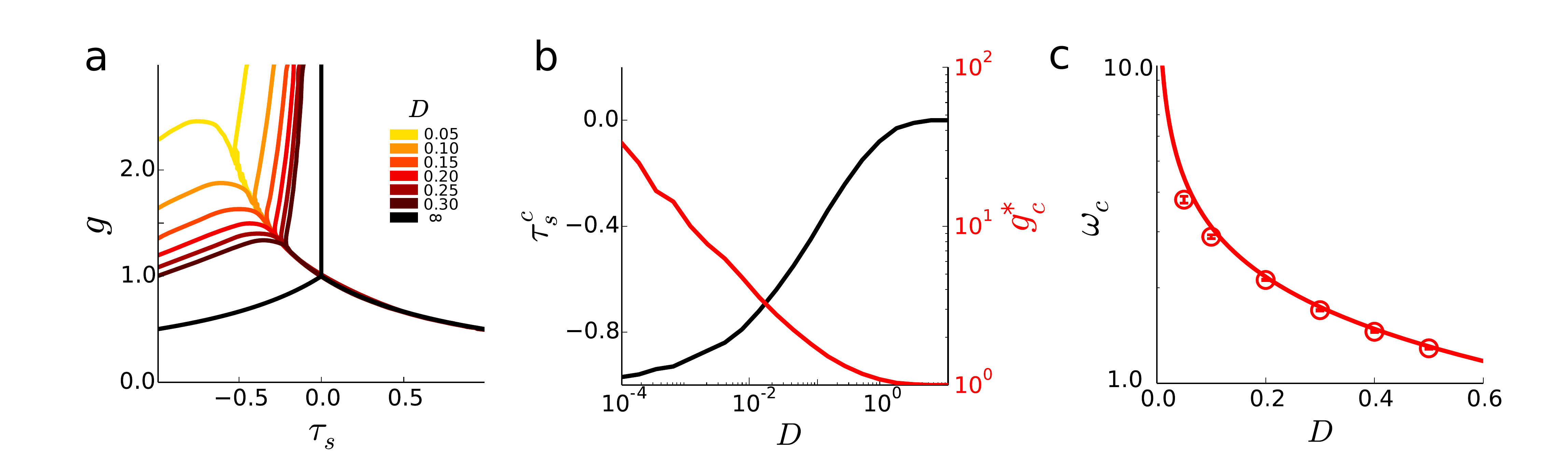}
  \caption{Effects of interaction delay on network dynamics. a. Bifurcation diagram for different values of interaction delay $D$. 
    b.  Values of $\tau_s^c$ and corresponding coupling $g_c^*$  as function of delay.   
    c. Bifurcation frequency  as a function of the normalized delay ($\tau_s=-0.7$, $g$ adjusted). Full line: theoretical prediction; red dots: simulation results (N = 500, n = 5, mean $\pm$ sem).}
  \label{Delay}
\end{figure}

The   critical  coupling  $g_c$   and  the  frequency
$\omega_c$ of the bifurcation strongly depend on the value of the
delay $D$ in  the interactions (Fig.~\ref{Delay}). As the value of  $D$ is increased from
$0$, for a given value of $\tau_s$  both $g_c$ and  $\omega_c$ progressively  decrease. 
At  the same
time,  $\tau_s^{c}$ increases, so  that the  oscillatory instability
occupies  an  increasing portion  of  phase  space.  In the  limit  of
infinitely long delays, $\tau_s^{c}$ tends to zero, so that the oscillatory
instability  extends  until  the  boundary between  antisymmetric  and
symmetric interactions.

\section{Heterogeneous Oscillatory State}

\begin{figure}
\includegraphics[width=1.0\textwidth]{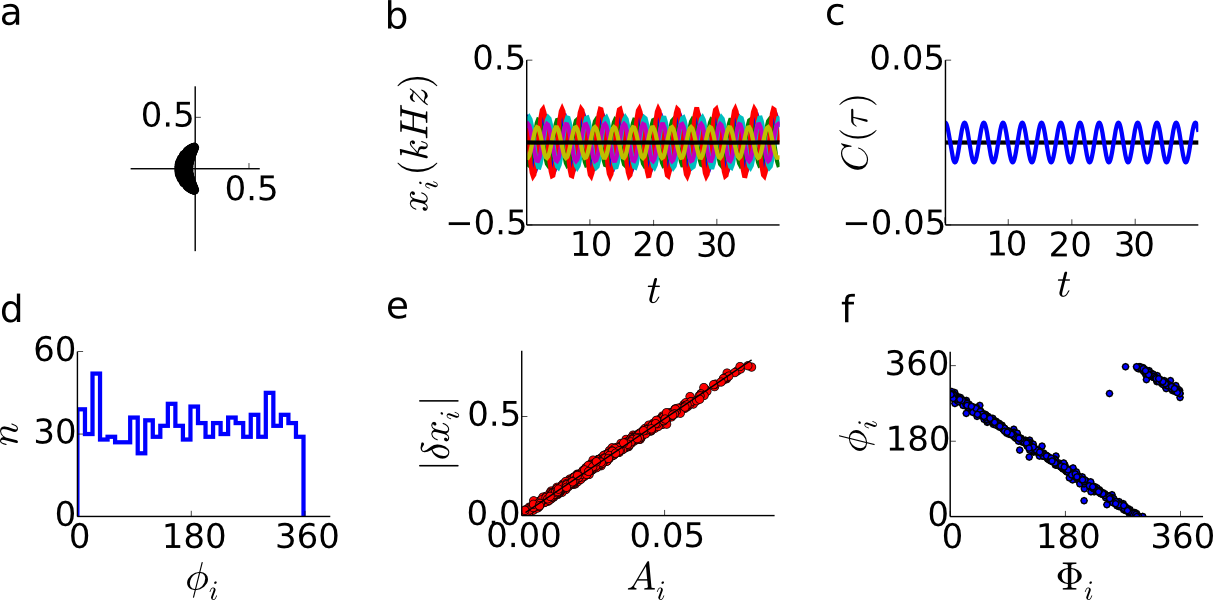}
\caption{Heterogeneous oscillations just above the instability ($g=1.4$).   Example simulation with only one $\lambda$ with a positive real part ($g=1.4$). The connection matrix was found so that only one eigenvalue was unstable. The simulations were performed using $\tau_s = -0.7$ and $D=0.2$ (N=1000).
    a. Distribution of the computed $\lambda$.
    b. Firing rate of 10 neurons superimposed.
    c. Mean auto-correlogram of all neurons $C(\tau)$ (blue trace) and auto-correlogram of the mean population activity $K(\tau)$ (black).
    d. Histograms of the phases of the firing rates of all neurons. 
    e. Correlation between the amplitudes of the firing rates and the module of each component of the principal eigenvector.
    f. Correlation between the phases of the firing rates and the phase of each component of the principal eigenvector. }
\label{oscillationsI}
\end{figure}

In the  region of  the oscillatory instability,  couplings larger  than $g_c$
lead  to a  novel  type  of heterogeneous  oscillatory  state that  we
investigated using numerical  simulations of Eq.~\ref{ref_eq}. In this
state, the different units appear  to oscillate with essentially identical frequencies, but random phases and
amplitudes (Fig.~\ref{oscillationsI}). To further characterize this state,  we defined the phase of each unit as the phase of the first peak ($t_{peak}$) with respect to an arbitrary reference common to all units ($t_0$)
\begin{equation}
\phi_i = (t_{peak}^i-t_0)\omega_0 \,\,\,\,\,\, \mathrm{modulo} \,\,\,\,\,\,\, 2\pi.
\end{equation} 
Correspondingly, the amplitude of each unit was defined as the peak-to-peak amplitude $|\delta x_i|$. The phases of different units appear to be uniformly distributed over $[0,2\pi]$ (Fig.~\ref{oscillationsI} d). 
In consequence  the average population activity is  constant in time, so that oscillations are apparent only on the level of individual units, but not on the population-average level. 
This was further confirmed by looking at the autocorrelation function of individual units, averaged over the network
\begin{equation}
C(\tau)=\frac{1}{N}\sum_{i=1}^{N} \int_{-\infty}^{+\infty} x_i(t)x_i(t-\tau) dt
\end{equation}

which showed clear oscillations. In contrast, the auto-correlation of network-averaged activity 
\begin{equation}
K(\tau)=\frac{1}{N^2}\int_{-\infty}^{+\infty} \sum_{i=1}^{N} x_i(t) \sum_{i=1}^{N} x_i(t-\tau)dt 
\end{equation}
was flat.

The structure of  the oscillatory activity can be  understood from the
linear analysis close  to the instability. In the  limit when a single
eigenvalue crosses the instability (Fig.~\ref{oscillationsI}a-f),  the oscillations are predicted to
occur along a single mode:
\begin{equation}
x_i(t)\sim e^{i \omega_c t}R_i^{(0)} + \mathrm{c.c.}
\end{equation}
where  $\{R_i^{(0)} \}_{i=1\ldots N}$  is the  right eigenvector  of $J$
corresponding  to  the unstable  eigenvalue  $\lambda_0=i\omega_c$. Within  this
limit, all  the units  share the same  oscillation frequency,  and the
distribution  of  amplitudes and  phases  of  the  different units  is
determined by the amplitudes and phases of the eigenvector $\{R_i^{(0)}=A_i e^{i\Phi_i} \}_{i=1\ldots
  N}$. A comparison with the simulations shows that close to the instability, 
  the phases and amplitudes of individual units are indeed perfectly predicted by the phases and amplitudes of the unstable eigenvector (Fig.~\ref{oscillationsI}e,f). The dynamics are therefore effectively one-dimensional, 
  and the random amplitudes 
  and phases  of the oscillation are a direct consequence 
  of the random distribution of the elements of the corresponding eigenvector ($A_i$ and $\Phi_i$ respectively).

\begin{figure}
\includegraphics[width=1.0\textwidth]{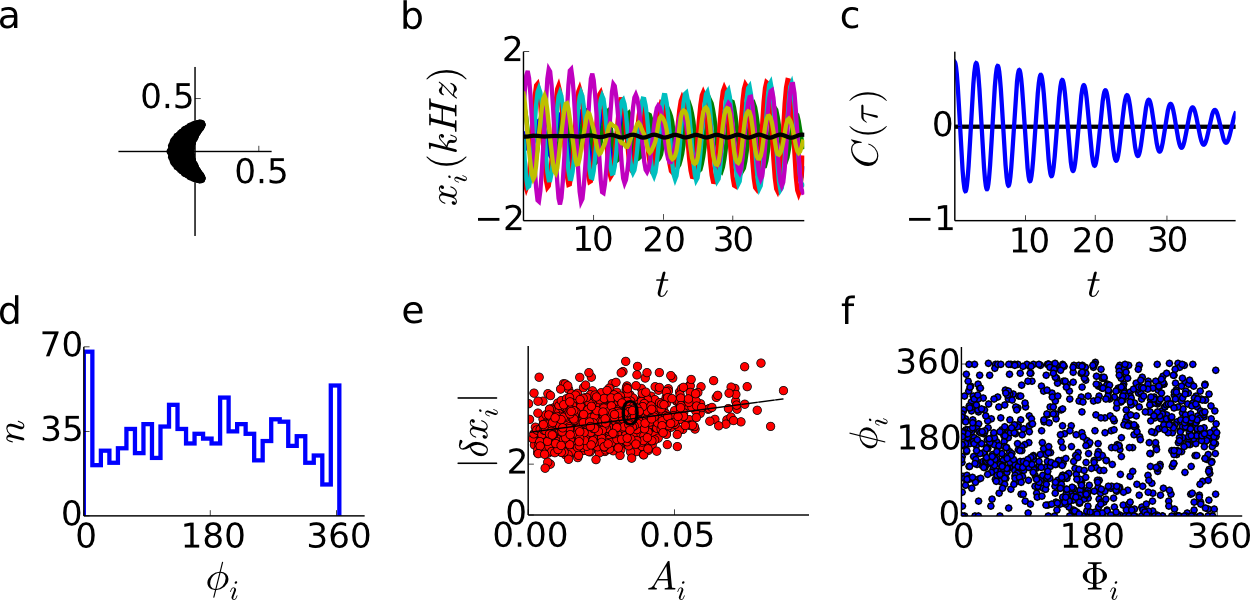}
\caption{Same as Fig.~\ref{oscillationsI}, but further above the instability ($g=2.0$), where several eigenvalues have crossed the instability line.}
\label{oscillationsII}
\end{figure}

As the coupling is further increased above $g_c$, an increasing number
of modes  become unstable  and the dynamics  become more  complex. The enveloppe of the
oscillations  starts to fluctuate (Fig.~\ref{oscillationsII}b), and  the
envelope of  the auto-correlation function of  individual units displays a slow decay (Fig.~\ref{oscillationsII}c). This
damping  of oscillations  may reveal  underlying chaotic  dynamics, an
issue we have not explored further. The individual  phases and amplitudes are not
anymore    accurately    predicted    by   the    first    eigenvector
(Fig.~\ref{oscillationsII}e,f), yet  the phases of  individual units remain
uniformly  distributed on  $[0,2\pi]$ (Fig.~\ref{oscillationsII}d), so  that the  summed population
activity is constant. The network therefore remains in a heterogeneous
oscillatory state,  in which individual units  strongly oscillate, but
this oscillation is not apparent on the population-average level.

\begin{figure}[h!]
  \includegraphics[width=0.8\textwidth]{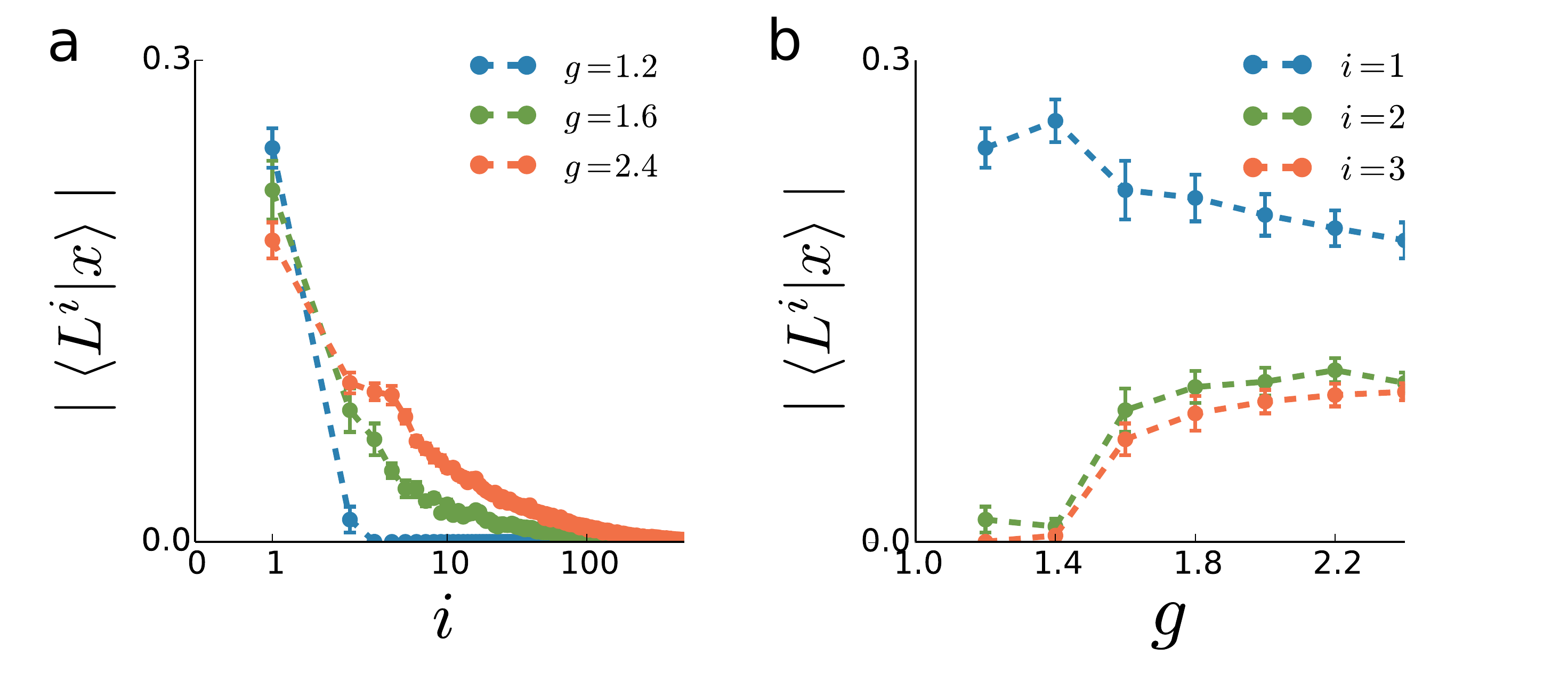}
  \caption{Dimensionality of the dynamics above the oscillatory instability. 
  a. Projection of network activity $x$ on the left eigenvector associated to the different eigenvalues, ordered by decreasing real part, 
  for increasing coupling $g$ (conjugate eigenvectors are not shown). 
  b. Projection of $x$ on the left eigenvector associated to the first, second or third eigenvalues as a function of $g$.    
  Parameter values: $\tau_s = -0.7$ and $D=0.2$, N=1000, n=30, mean $\pm$ sem.}
  \label{dimensions}
\end{figure}

To  further  quantify  to  dimensionality  of the  dynamics,  we  have
computed projections of  the activity on the left  eigenvectors of the
coupling matrix, defined  as
\begin{equation}
 \frac{1}{T}\sum_{t=1}^{T}\langle L^{i} | x_t \rangle / \| x_t \|.
\end{equation}
with $T$ being the length of the time-window used for all computations ($T=10^7$).
As the coupling
is increased,  the number of non-zero projections  increases. This can
be  seen both  when  looking at  all  the sorted  modes  with a  fixed
coupling (Fig.~\ref{dimensions}a) or varying  the coupling for a fixed
eigenmode  (Fig.~\ref{dimensions}b).  The  oscillatory  state explores
more  and more  dimensions  as  the coupling  is  increased above  the
instability.

\section{Discussion}

In  summary, we have  described a  heterogeneous oscillatory  state in
randomly   coupled   networks  units   with   delayed  and   partially
anti-symmetric interactions.  This oscillatory  state is governed by a
novel type of bifurcation, in which a continuum of eigenvalues crosses
the  instability  line  at  a  non-zero  frequency.   In  this  state,
individual  units oscillate  with random  phases, so  that  the summed
population activity remains constant  in time.  

From this perspective, the  heterogeneous oscillatory state bears some
similarity with the splay-state, an asynchronous regime in networks of
interacting oscillators  \cite{Abbott-van-Vreswijk1993, tattini}.  One
important  difference  is  that  in  the  classical  splay  state  the
individual units  oscillate even in absence of  coupling. In contrast,
in the network considered here, the individual units are not intrinsic
oscillators.  In  absence of coupling,  their activity is  constant in
time,  and the oscillations  appear only  when the  coupling increases
beyond a critical value. Note also that while both the splay state and
the heterogeneous  oscillatory state  are ``asynchronous" in  the sense
that  the  macroscopic activity  displays  no  oscillations, they  are
distinct  from other  types of  asynchronous  states \cite{brunel2000,
  renart, Ostojic} in which the individual units do not oscillate.

The  instability described  here is  also related  to  the oscillatory
Turing  instability  described in  the  context  of random  ecological
networks  \cite{Hata}.  That  bifurcation   is  also  generated  by  a
continuous  part of  the eigenspectrum  of the  stability  matrix, and
leads to  heterogeneous oscillations in  which different nodes  in the
network oscillate  out of phase.  The underlying mechanism  is however
different  as in Ref.~\cite{Hata} the  interactions between  nodes are  symmetric  and not
delayed.

Following previous works, we considered here a simplified network with
a  Gaussian connectivity  matrix,  and a  transfer function  symmetric
around  zero.   Such a  model  does  not  implement basic  biophysical
constraints  such  as segregation  between  excitatory and  inhibitory
neurons,  or the  requirement that  neural activity  is  positive. The
influence  of such  constraints on  the zero-frequency  instability to
rate  chaos have  been  investigated  recently in  a  series of  works
\cite{Ostojic,Kadmon,Harish,Mastroguiseppe}.  These studies focused on
the  asymmetric case.  Determining  how the  heterogeneous oscillatory
state described  here occurs in  more realistic networks  will require
combining   excitation-inhibition   segregation   with  symmetry   (or
asymmetry)  in  the   connections.   This  represents  an  additional
challenge,  as   both  ingredients  influence  the   spectrum  of  the
interaction   matrix.  Note   that   reciprocal  connections   between
excitatory  and inhibitory  cells may  represent a  natural  source of
anti-symmetric connectivity. Such  connections have been in particular
reported in the  olfactory bulb, and linked to  the gamma oscillations
occurring  there \cite{Li}.  It  would be  interesting to  investigate
further  whether  these oscillations  are  related  to the  phenomenon
described here.

Oscillatory  activity is  widespread  in biological  networks, and  in
particular  in the  brain  \cite{buzsaki}. Possible  origins of  these
oscillations have mainly been  conceptualized in terms of bifurcations
in  low-dimensional  dynamical  systems.  Together with  other  recent
results   \cite{delMolino-Pakdaman2013},  this   study   suggest  that
high-dimensional,  disordered networks can  lead to  novel oscillatory
mechanisms  that  should  be  taken  into  account  when  interpreting
biological data.

\section*{Acknowledgement}

We are grateful to Vincent Hakim  for discussions. This work was funded by the Programme Emergences of City of Paris, and the program
``Investissements d'Avenir'' launched by the French Government and
implemented by the ANR, with the references ANR-10-LABX-0087 IEC and
ANR-11-IDEX-0001-02 PSL* Research University. 
  
\end{document}